%Paper: hep-ph/9405392
%From: RONCAGLI@ind.physics.indiana.edu
%Date: 26 May 94 17:11:00 EST

\normalbaselineskip=12pt
\baselineskip=12pt
%\magnification=\magstep1
\hsize 15.0truecm \hoffset 0.7 truecm
\vsize 22.0truecm
%May 19, 1994

\nopagenumbers
\headline={\ifnum \pageno=1
\hfil \else\hss\tenrm\folio\hss\fi}
\pageno=1

\def\lsim{\mathrel{\rlap{\lower4pt\hbox{\hskip1pt$\sim$}}
    \raise1pt\hbox{$<$}}}     %less than or approx. symbol
\def\gsim{\mathrel{\rlap{\lower4pt\hbox{\hskip1pt$\sim$}}
    \raise1pt\hbox{$>$}}}   %greater than or approx. symbol

%%%%%% (defining the references) %%%%%%%

\def\QR{1}         % Quigg, Rosner
\def\LSG{2}        % Lucha Schoberl Gromes
\def\MNSMM{3}      % Mukherjee et al.
\def\Ri{4}         % Richard
\def\DBL{5}        % DBL, Int.J.Mod.P.
\def\Di{6}         % Dieckmann
\def\FS{7}         % Flamm Schoberl
\def\GR{8}	   % Gasiorowicz Rosner
\def\KR{9}         % Kwong, Rosner
\def\CL{10}        % Cohen Lipkin
\def\Fe{11}        % Feynman
\def\He{12}        % Hellmann
\def\GM{13}        % Grosse Martin
\def\Song{14}      % Song
\def\WL{15}        % Wang, DBL
\def\don{16}       % DBL, PRD 40, 4196 (1989)
\def\Ei{17}        % Eichten et al.
\def\GI{18}        % Godfey Isgur
\def\BBHM{19}      % Blask et al.
\def\DL{20}        % DBL, PRD 40, 3675 (1989)
\def\BO{21}        % Bertlmann Ono
\def\PDG{22}       % Particle Data Group 1992
\def\CDF{23}       % CDF collaboration, B_s measurement
\def\ALEPH{24}     % ALEPH collaboration, B_s measurement
\def\Ono{25}       % Ono
\def\BCN{26}       % Bhadhuri et al.
\def\SB{27}        % Silvestre-Brac
\def\Fu{28}        % Fulcher
\def\M{29}	   % Martin 1980
\def\Ma{30}        % Martin
\def\BM{31}        % Bertlmann Martin
\def\Th{32}	   % Thirring
\def\QEQ{33}       % Quigg; Eichten and Quigg
\def\Ba{34}        % Bagan et al.
\def\JM{35}	   % Jain and Munczek
\def\Quigg{36}     % Quigg, private comm.
\def\UA1{37}       % UA1 measurement of Lambda_b
\def\OPAL{38}      % Opal measurement of Lambda_b

\hfill IUHET 278

\line{\hfil May, 1994}

\bigskip
\vskip 12pt
\centerline{\bf Predicting  the masses of heavy hadrons
without an explicit Hamiltonian}
\vskip 36pt
\centerline{R. Roncaglia, A. Dzierba, D. B. Lichtenberg,
and E. Predazzi,\footnote{$^{*}$}{On leave
from the University of Torino, Italy}}

\centerline{\it Department of Physics,
Indiana University, Bloomington, Indiana, 47405}
%, USA}

\vskip 1.0in

\item{}
There are striking regularities in the
masses and mass differences of known hadrons.
Some of these regularities can be understood from
known general properties of
the interactions of quarks without a need to specify
the explicit form of the Hamiltonian. The Feynman--Hellmann
theorem is one of the tools providing this
understanding.  If the  mass regularities are exploited,
predictions can be made of the masses of as yet
undiscovered hadrons. In particular, it is found
that the mass of the $B_c^*$ is $6320\pm 20$ MeV.
Predictions concerning i) excited vector mesons,
ii) pseudoscalar mesons, iii) $P$-wave mesons,
and iv) ground-state spin 1/2 and 3/2 baryons are also made.

\bigskip
\item{} PACS numbers: 12.15.Ff, 12.40.Yx, 14.20.-c, 14.40.-n
\vfill \eject

\centerline{I. INTRODUCTION}
\medskip

Thus far, the quark potential model has been the
most successful tool enabling physicists to calculate
the masses of normal mesons and baryons
containing heavy quarks. We call attention
to several reviews on the subject [\QR --\GR].
However, potential models suffer from the fact
that, although motivated from QCD, so far, they cannot
be derived from that theory.
In this paper we make predictions about hadron masses
with complementary methods
which use general properties of the potential (or,
more generally, of the interaction) but
not its specific form. Among the complementary
methods are:
\item{(1)}
Obtaining constraints on hadron (and quark)
masses [\QR,\KR,\CL]
from the Feyn\-man--Hellmann theorem [\Fe,\He];
\item{(2)} Using theorems which relate the ordering of
bound-state energy levels to
certain properties of potentials [\GM];  and
\item{(3)} Taking advantage of regularities
in known hadron masses to obtain estimates of as
yet undiscovered hadrons using either interpolation [\KR] or
semi-empirical mass formulas [\Song,\WL].

We exploit these methods
to obtain constraints on quark and hadron
masses.  We also  provide new theoretical justification
for the methods we use
and make predictions for the masses of
as yet undiscovered mesons and baryons.

We devote
considerable effort to making what we believe is
a good prediction for the mass of the $B_c^*$ vector
meson, using the Feynman--Hellmann theorem. We also
discuss in some detail the ground-state
pseudoscalar mesons and the ground-state baryons
(both spin 3/2 and spin 1/2), because additional issues
arise in these cases. We discuss only briefly the
excited vector mesons and $P$-wave mesons (tensors,
axial vectors, and scalars), although we give
some predictions in these cases as well.
\medskip

\centerline{II. THE FEYNMAN--HELLMANN THEOREM}
\medskip
Some years ago, Feynman [\Fe] and Hellmann [\He]
independently
showed that if a Hamiltonian $H$ depends on a parameter
$\lambda$, then the bound-state energy eigenvalues
$E(\lambda)$ vary with $\lambda$ according to the
formula
$$\partial E / \partial \lambda=
\langle \partial H /  \partial \lambda  \rangle,
\eqno(1)$$
where the expectation value
is taken with respect to the normalized eigenfunction
belonging to $E$.
The Feynman--Hellmann theorem was applied to
quarkonium physics [\QR,\CL], with $\lambda=\mu$,
the reduced mass of the system.
For example, Quigg and Rosner [\QR]
applied the theorem to the  nonrelativistic
Hamiltonian $H=p^2/(2\mu) +V$, where
$V$ is an interaction which is
assumed to be flavor independent, and
therefore independent of $\mu$.  Then
$$\partial E / \partial \mu=
-\langle p^2\rangle/ (2\mu^2)  <0, \eqno(2)$$
{\it i.e.}, $E$ decreases monotonically as $\mu$
increases because $p^2$ is a positive definite operator.
Of course, if $V$ depends on $\mu$ but
$\langle\partial V/\partial\mu\rangle\leq 0$, then
$\partial E/\partial\mu < 0$ still remains valid.

Even in the case of some many-body Hamiltonians with
relativistic kinematics, the
Feynman--Hellmann theorem may be applied to give
useful information about how eigenenergies change when
constituent masses $m_i$ change [\don].
As an example, we consider a Hamiltonian $H$, given by
$$H= \sum_i[({\bf p}_i^2+m_i^2)^{1/2}-m_i] +
V({\bf r}_1,...{\bf r}_n;m_1,...m_n),
\eqno(3)$$
where we have let the interaction $V$ depend explicitly
on the $m_i$ ($i=1,2...n$). Taking the partial derivative
with respect to $m_i$  and using (1), we obtain
$$\partial E / \partial m_i=
\langle m_i / (p_i^2+m_i^2)^{1/2}\rangle -1
+\langle \partial V /\partial m_i\rangle. \eqno(4)$$
We can see from Eq.\ (4) that if
$$\langle\partial V /\partial m_i \rangle \leq 0, \eqno(5)$$
then,
$$\partial E / \partial m_i<0, \quad m_i=1,2...n. \eqno(6)$$
These inequalities are a generalization of (2).

In the remainder of this section, we restrict
ourselves to the case in which (5) and (6) both hold,
so that an increase in one or more $m_i$ leads to a
decrease in $E$.
We generalize the definition of $\mu$ to be
$$\mu^{-1} = \sum  m_i^{-1}.\eqno(7)$$
We now note that an increase in one or more $m_i$ results
in an increase in $\mu$ as well as a decrease in $E$.
{\it Under these circumstances}, a change
$\delta \mu$ results in a change
$\delta E$ in the opposite direction.
Therefore, $E$ will be monotonically decreasing with
increasing $\mu$, provided the
increase in $\mu$ arises from an increase in one
or more $m_i$.
We can state this result in the form
$$\delta E/\delta \mu<0, \eqno(8)$$
{\it if the changes in
the $m_i$ are all in the same direction}.
The inequality (8) turns out to be
a powerful tool for obtaining constraints on
quark and  hadron masses. As we shall see in Sec.\ V,
this inequality holds empirically for vector mesons
even in the absence of the restriction that all $m_i$
change in the same direction.

For a two-body Hamiltonian of the form (3) with
a flavor-independent potential, we can say something
more. Consider $E$ as a function of $\mu$ and ${\cal M}$,
where ${\cal M}=m_1+m_2$.
We can then show explicitly that
$$\partial E/\partial \mu<0,\quad
\partial E/\partial {\cal M}<0.\eqno(9)$$
It follows that
even if $m_1$ and $m_2$ change
in opposite directions,
if  $\mu$ increases and ${\cal M}$ either remains constant
or increases, then $E$ decreases. This result
holds even in the presence of some flavor-dependent
interactions, including a colormagnetic interaction for
vector mesons, as we shall see in the next section.

\medskip
\centerline{III. APPLICATION TO MESON AND BARYON
EIGENENERGIES}
\medskip
We adopt a constituent quark picture, assuming that
a meson is composed of a quark and an antiquark, and
a baryon is composed of three quarks. We confine ourselves
to hadrons containing $u$, $d$, $s$, $c$, and $b$ quarks.
Furthermore, we neglect any violation of isospin
and let $m_u=m_d=m_q$.
As usual, we assume that the quark masses $m_i$ satisfy the
inequalities
$$m_q<m_s<m_c<m_b. \eqno(10)$$

In order to apply (6) to hadrons, we need to discuss
for which hadrons (5) is likely to hold.
The interaction $V$
can be written as $V_0+ V'$, where $V_0$ is independent
of quark flavors and $V'$ depends on flavor.
The term $V_0$ is the static quark-antiquark
potential, which is commonly assumed [\Ei]
to contain a Coulomb-like term, an
approximately linear confining term,
and  a constant term, all independent of flavor.
In the Fermi--Breit approximation, $V'$ contains
both  spin-dependent and spin-independent terms which
are explicitly functions of flavor through
the quark masses. However, most phenomenological
treatments of quarkonia have not needed the Fermi--Breit
spin-independent term [\DBL], and we neglect it here.
In states with zero orbital angular momentum, the
expectation values of the tensor and spin-orbit terms
of the Fermi--Breit interaction vanish, leaving the
colormagnetic interaction as the only spin-dependent term.
We write the colormagnetic term $V_{cm}$ in the form [\GI]
$$V_{cm}=- \sum_{i<j}\lambda_i\cdot \lambda_j
f(r_{ij})\sigma_i\cdot \sigma_j/(m_im_j),
\eqno(11)$$
where $\sigma_i$, $\sigma_j$ are Pauli spin matrices,
$\lambda_i$, $\lambda_j$ are Gell-Mann SU(3) matrices,
and $f(r_{ij})$ ($r_{ij}=|{\bf r}_i-{\bf r}_j|$) are
positive definite operators.

If $V=V_0 + V_{cm}$, we obtain
$$ \langle \partial V/\partial m_i\rangle =
\sum_{j\not=i} \langle f(r_{ij}) \rangle
\langle\lambda_i\cdot \lambda_j\rangle
\langle\sigma_i\cdot \sigma_j\rangle/(m_i^2m_j).
\eqno(12)$$
The quantity
$\langle\lambda_i\cdot \lambda_j\rangle$ is
negative for a quark-antiquark
pair in a meson ($-16/3$) and for all quark
pairs in a baryon ($-8/3$). Also,
$\langle\sigma_i \cdot \sigma_j\rangle$
is $-3$ if  two quarks (or a quark and antiquark)  are in
a spin-zero state,  and 1 if they are in a spin-one state.
Now in vector mesons and spin-3/2 baryons, we have
$\langle\sigma_i\cdot \sigma_j\rangle=1$ for
all quark pairs. Then
we see from (12)  that  the eigenenergies of these hadrons
satisfy (5). We therefore expect the energy eigenvalues
of vector mesons  and spin-3/2 baryons to satisfy (5), and
therefore (6) as well. In the two-body case (vector
mesons), (11) can be written
$$V_{cm}= (16/3) f(r_{12})/(\mu{\cal M}).  \eqno(13)$$
We can see explicitly from (13) that
$\langle \partial V/\partial \mu \rangle<0$,
$\langle \partial V/\partial {\cal M}\rangle< 0$,
so that (9) holds, as we stated in the previous section.

For pseudoscalar mesons the sign of the colormagnetic
term is negative, and
the interaction violates (5). For spin 1/2 baryons,
the three colormagnetic terms in (11)
are either all negative or one term is positive and
two are negative, so that (5) is sometimes violated.
Thus, we expect that
pseudoscalar mesons and spin-1/2 baryons might violate
(6) for small $m_i$, where the contribution
from (12) is large and positive.

In addition to $V_{cm}$, terms arising from instantons may
contribute to $V'$. These terms are apparently important in
states in which two quarks (or a quark and an antiquark)
have spin and orbital angular momentum zero
(pseudoscalar mesons and spin-1/2 baryons) [\BBHM].
Instantons tend to mix the wave functions of
certain mesons, like the $\eta$ and $\eta'$ (which
contain both $q\bar q$ and $s \bar s$ in their wave
functions, and perhaps some glueball admixture as well).
Such states are unsuitable for
our scheme, as, in order to compute the reduced mass of
a system, we must know its quark content.

If we confine ourselves to mesons containing $q(=u,d)$,
$s$, $c$, and $b$ quarks, then
10 different ground-state vector mesons  and
20 different ground-state baryons of spin 3/2 exist.
Of the former, 9 are experimentally known; of the latter,
only 4 are known, none of which contains a heavy quark.

Let a particular energy eigenvalue of a meson
containing a quark $i$ and an antiquark $j$ be
$E_{ij}$, where we suppress an index labeling which
eigenvalue we are referring to.
Likewise, we denote an energy eigenvalue of a
baryon by $E_{ijk}$. If we replace a single quark
by a heavier quark, then, because of
(6), the eigenenergy of the new hadron will be smaller
than that of the old one.  By continuing this process,
we obtain chains of inequalities among the $E_{ij}$
and also among the $E_{ijk}$.
We consider the {\it longest chains} of inequalities
for which (6) holds.  A longest chain arises when
we start with the lightest hadron (containing only
$q$-type quarks) and replace each of its quarks
one at a time
by the next heavier quark, each time obtaining a different
hadron (i.e., we replace $q$ by $s$,
$s$ by $c$, and $c$ by $b$).
A longest meson chain contains 7 eigenenergies,
and a longest baryon chain contains 10.

There are five different
longest chains for mesons, all of which have
the same first two and last two eigenenergies.
At least one of the three intermediate eigenenergies
differs in each of the five chains.
One longest chain for mesons is
$$E_{bb}<E_{bc}<E_{cc}<E_{cs}
<E_{ss}<E_{sq}<E_{qq}. \eqno(14)$$
The four other longest chains have intermediate
eigenenergies
$$E_{cc},E_{cs}, E_{cq}; \quad
E_{bs},E_{cs},E_{ss}; \quad
E_{bs},E_{cs}, E_{cq}; \quad
E_{bs},E_{bq},E_{cq}.\eqno(15)$$
It follows  that
(6) holds for the mesons with eigenenergies
as in (14) or (15), whereas this may not necessarily
be the case when one quark mass increases
and the other decreases.

We next turn to the spin-3/2 baryons, and denote their
energy eigenvalues by $E_{ijk}$. From our discussion
we can write down a number of longest chains. One
such chain is
$$E_{bbb}<E_{bbc}<E_{bcc}<E_{bcs}<E_{bss}<E_{css}<E_{sss}
<E_{ssq}<E_{sqq}<E_{qqq}. \eqno(16).$$
All longest distinct baryon chains, of which
we have counted
42,  contain the same first two and last two
eigenenergies.
\medskip

\centerline{IV. CONSTRAINTS ON QUARK MASS DIFFERENCES}
\medskip
The energy eigenvalues $E_{ij}$ and $E_{ijk}$ do not
include the quark rest energies. Therefore, the mass $M$
of a hadron is given by
$$M= E +\sum m_i, \eqno(17)$$
where we have suppressed the indices on $E$ and $M$.
As has been pointed out by a number of authors
[\QR,\LSG,\DL,\BO]
Eq.\ (17) can be used together with (8) to obtain
constraints on quark mass differences in the
form of inequalities. What is new
in our treatment is our justification of the use
of (8) for longest chains of vector meson and spin-3/2
baryon eigenenergies. In particular, this means
that we do not need to use as input spin-averaged
hadron masses, which are normally calculated in a
perturbative approximation.

In order to obtain inequalities among quark masses,
we use the experimental values of hadron masses,
including the $\rho$, $K^*$, $\phi$, $D^*$, $D_s^*$, $B^*$,
$B_s^*$, $J/\psi$, and $\Upsilon$ mesons (of the
10 vector mesons, only the $B_c^*$ is missing) and the
$\Delta$, $\Sigma^*$, $\Xi^*$, and $\Omega$ baryons
(only 4 of the 20 spin-3/2 baryons are known).
When the isospin of a state is greater than zero, we
average over the members of the isospin multiplet.
For the $\bar qq$ state,
we believe it is better to choose the $\rho$ than the
$\omega$ because the latter
might have a small admixture of $\bar ss$, whereas
such an admixture in the $\rho$ violates isospin.  Note that
in our scheme the $\omega$ is degenerate with the $\rho$,
although it is actually 15 MeV heavier.
In fact, in not distinguishing between $u$ and $d$ quarks,
we consider
the $\omega$ and $\rho$ together as just a single one of
the 10 distinct vector mesons in our scheme, which is
concerned only with masses. We have given an {\it a priori}
reason for choosing the $\rho$ instead of the $\omega$.
In the next paragraph we give an {\it a posteriori} reason.
However, if instead of choosing the $\rho$, we choose
the $\omega$, our predictions are not appreciably affected.
All the meson masses are taken from the Particle Data Group
[\PDG], except the mass of the $B_s^*$, which comes from two
recent measurements [\CDF,\ALEPH] of the mass of the $B_s$
plus a measurement of $m(B_s^*)-m(B_s)$ (which needs
confirmation) quoted in [\PDG].

Using the masses of the observed ground-state vector mesons,
we obtain the following inequalities:
$$m_s-m_q>M(K^*)-M(\rho)=126\pm 4 \ {\rm MeV},$$
$$m_c-m_s>M(D^*)-M(K^*)=1115\pm 4 \ {\rm MeV},\eqno(18)$$
$$m_b-m_c>M(B^*)-M(D^*)=3316\pm 6 \ {\rm MeV}.$$
The errors are partly experimental and partly due to
our assumption of isospin invariance.
If we substitute the $\omega$ for the $\rho$, the
right hand side of the first of these inequalities becomes
smaller by 15 MeV. However, we can regain the stronger
inequality (larger right-hand side) by considering
$$m_s-m_q >M(\phi)-M(K^*)= 125\pm 4\ {\rm MeV}.\eqno(19)$$
This fact gives an {\it a posteriori} justification of our
decision to use the $\rho$, rather than the $\omega$,
in our set of vector mesons.

One can obtain apparently stronger inequalities
by using the spin-averaged values for vector and
pseudoscalar mesons. The results in MeV are [\DL]
$$m_s-m_q>184\pm 4,\quad m_c-m_s> 1180\pm 4, \quad
m_b-m_c> 3343\pm 4. \eqno(20)$$
However, the spin averaging process relies
on a perturbative treatment of the spin-dependent
forces, which may not be justified for light mesons.
In fact, we shall see in Sec.\ VI
that when we vary the quark masses
to obtain a best fit to the data, the quark mass differences
violate (20) but satisfy (18).
Still other authors [\BO] have obtained inequalities among
quark masses using a variety of assumptions, but not
if the Hamiltonian contains both relativistic
kinematics and a flavor-dependent interaction.

As we have remarked, only 4 ground-state baryons of
spin 3/2 have been observed thus far. These
lead only to inequalities for $m_s-m_q$. The
strongest of these is
$$m_s-m_q>M(\Sigma^*)-M(\Delta)=153\pm 4 \ {\rm MeV}.
\eqno(21)$$
This inequality is stronger than the corresponding
inequality we obtained from mesons.
As we shall see in Sec.\ VIII, in order to obtain
a best fit to the baryon data, we must use
quark mass differences for baryons which are
up to 35 MeV larger than the corresponding quark
mass differences for mesons.
\medskip

\centerline{V. SATISFYING THE QUARK MASS CONSTRAINTS}
\medskip
Many different sets of quark constituent masses are used
in the literature, most of them obtained
from fits to spectroscopic data. We show in Table I
a selection of these sets
[\FS,\KR,\Song,\WL,\Ei,\GI, \BO,\Ono--\Fu],
including in the first row the set that we use
in this work for vector mesons. (In the next section,
we explain how we arrive at this
mass set.) We can see from Table I
that the sets of masses given in the first seven rows
satisfy the inequalities given in (18), whereas the
sets of masses given in the last five rows violate
one or more of these inequalities. Therefore, if one
calculates the vector meson masses with
any one of the last five sets and
a potential which is flavor independent
except for a conventional colormagnetic term,
one will obtain results in disagreement with experiment.
This disagreement will occur independently of whether
one uses a nonrelativistic Schr\"odinger equation
or a  wave equation with relativistic kinematics of
the form given in Eq.\ (3).

We use the {\it experimental
values} of the vector meson masses together with the
input values of the quark masses to calculate the meson
eigenenergies $E_{ij}$ from Eq.\ (17). We can plot
these eigenenergies as a function of $\mu$.
For illustrative purposes, we use four different sets of
input quark masses from Table I,
and show in Fig.\ 1 how $E$ varies as a function
of $\mu$ for the ground-state vector mesons.
For completeness, we include mesons
containing  $q\bar b$, $q\bar c$, and $s\bar b$ as well as
$s\bar c$, $s\bar s$, and $c\bar c$ in order to see
how $E$ varies with $\mu$ when  $m_1$ increases and
$m_2$ decreases.
In Fig.\ 1 (a) and (b), the quark masses satisfy
all the inequalities (18), whereas in Fig.\ 1 (c) and
(d), the quark masses violate at least one of these
inequalities.

We see from Fig.\ 1 (a) and (b) that, when the quark
masses satisfy (18), $E$ appears indeed to be
a monotonically decreasing function of $\mu$ for the
vector mesons. From Fig. 1 (c) and (d) we see that, with
quark masses violating (18), $E$, as obtained from
the {\it observed meson masses},
is not monotonically decreasing as a function of $\mu$.
The sets of quark masses used in (c) and (d)
seem {\it a priori} as reasonable as those of
sets (a) and (b). However, if we use the quark
masses of (c) or (d) with a flavor-independent potential
and a conventional colormagnetic term,
the {\it calculated} values of $E$ must
be monotonically decreasing and therefore
in disagreement with experiment.

We see from Fig.\ 1 (a) and (b) that, not only is $E$
monotonically decreasing as a function of $\mu$, but is
also concave upward.  We cannot  prove
concavity from the Feynman--Hellmann theorem. However,
in the two-body nonrelativistic case, we can show
for a power-law potential  of the form
$$V=\alpha r^{\beta},\quad \alpha\beta >0,\quad
\beta>-2 \eqno(22)$$
that the concavity condition
$$\partial^2 E/ \partial \mu^2 >0 \eqno(23)$$
is true. This result follows directly from the scaling
property of the Schr\"odinger equation with a power-law
potential [1]. The fact that (23) holds for a power-law
potential is relevant because, as has been emphasized
by Martin [\M,\Ma], the quarkonium static
potential can be well approximated by (22).
If we include a Fermi--Breit term in the potential of form
given by Eq.\ (12), (23) remains valid. Of course,
we are interested in the curvature of $E$ as a function
of $\mu$ without concerning ourselves with how ${\cal M}$
varies. In the presence of
the colormagnetic interaction (11), we can say that the
curvature is positive, provided that as $\mu$ increases,
${\cal M}$ does not decrease. It has been
shown [\BM, \Th] that in the two-body
nonrelativistic case with a flavor independent potential,
$$\partial^2 E/ \partial \lambda^2 <0, \eqno(24)$$
where $\lambda = 1/\mu.$
Not only is $E$ concave upward as a function of $\mu$ in
a class of nonrelativistic two-body models, but
it turns out that
$$\delta^2 E/\delta \mu^2 >0 \eqno(25)$$
holds empirically for vector mesons (see Fig.\ 1 a and b)
and also for spin-3/2 baryons.
\medskip

\centerline{VI. PREDICTING THE $B_c^*$  MASS}
\medskip
Of the 10 vector mesons in our scheme, only the $B_c^*$
has not yet been seen.
We can use the inequalities (8) and (25) together
with the experimental vector meson masses and Eq.\ (17)
to estimate the mass of the $B_c^*$ by interpolation.
We do this by assuming that $E(\mu)$ can be approximated
by a simple curve containing only a few parameters.
The simplest curve (containing only 2 parameters) is
a straight line, but a straight line violates the
concavity condition (25). We therefore  approximate
$E(\mu)$ with a three-parameter curve. We emphasize that
the functional form of the curve has no theoretical
significance other than that it satisfies  the inequalities
(8) and (25). In addition to the three parameters of the
curve, we have four additional parameters, namely, the
quark masses $m_q$, $m_s$, $m_c$, and $m_b$. Using a given
three-parameter curve, we vary the 7 parameters
in order to obtain a best fit to the eigenenergies
$E_{ij}$. We then obtain the value of $E_{bc}$ from the
fitted curve and use the fitted masses $m_c$ and $m_b$
to obtain $M(B_c^*)$.

We at once encounter a difficulty in our scheme: namely,
that a longest chain of meson eigenenergies contains
only 7 members, and one of these ($E_{bc}$) is
unknown. Therefore, if we use a longest chain,
we have 7 parameters and only 6 data points
so that the parameters are not uniquely determined.
We overcome this difficulty by using all 9 known meson
eigenenergies, assuming that (8) and (25) hold even in
this case. We find that our assumption is consistent
with experiment:
namely, we can find a set of quark masses such that
the meson eigenenergies, as calculated from the
experimental values of the meson masses with the aid of
(17), satisfy (8) and (25).
The procedure of including all nine known meson masses
constrains the parameters much more than using only six
masses. However, even when we use 9 data points, it turns
out that the quark mass differences are much more
constrained than the quark masses themselves.

Kwong and Rosner [\KR] previously used the interpolation
method with two three-parameter curves (quadratic
and Pad\'e), although without any theoretical justification
of the inequalities (9) and (25). We have used three
different three-parameter curves, an exponential
$$E= a\exp(-\mu/b)-c,  \eqno(26)$$
a quadratic
$$E=a +b\mu +c\mu^2, \eqno(27)$$
and a hyperbolic (or Pad\'e)
$$E= a/(\mu+b)-c, \eqno(28)$$
where $a$, $b$, and $c$ are parameters to be varied.
In principle, these parameters are functions
of the sum of the quark masses as well as a function
of $\mu$, but from our previous discussion, we expect $E$
to be decreasing when plotted
as a function of $\mu$ with $a$, $b$
and $c$ independent of any other
masses in the problem.  We obtain comparable
fits to the data with all three curves, and the
meson energies are quite stable to our choice of functions.

We show in Fig.\ 2 our fit to the vector
meson energies, with an exponential curve as an example,
and the set of quark masses given in the first row of
Table I.  These quark masses are rounded to the nearest 10 MeV
and are based on a somewhat arbitrary
choice of 300 MeV for the mass of the $u$ and $d$ quarks.
We can get comparable fits to the data
for  quark masses which differ from our choices by 100 MeV
or more, but the mass differences are much more constrained.
With the exponential fit,
the values of the parameters with our quark masses are
$$a= 754\ {\rm MeV}, \quad b= 1375\ {\rm MeV}, \quad
c= 506\ {\rm MeV}. \eqno(29)$$

Our result for the $B_c^*$ mass is
$$M(B_c^*)=6320\pm 20 \ {\rm MeV}. \eqno(30)$$
We have estimated the theoretical error partly from
the spread in values obtained using
the different functional forms for $E(\mu)$ given in
Eqs.\ (26--28) and partly from values obtained
with different longest chains and various constraints
on the quark masses. Our quoted errors in Eq.\ (20) and
our subsequent predictions reflect the stability inherent
in our procedures. Our predicted value of the $B_c^*$ mass
is given in Table II in the first row of column 4.

Our value of $M(B_c^*)$
is more stable to the choice of curve than
the result of Kwong and Rosner [\KR]:
$6284 <M(B_c^*)< 6349$ MeV.  Perhaps one reason for this
is that we differ from those authors
in the choice of the function $\chi^2$
to be minimized. We choose
$$\chi^2=\sum [E(\mu)-E({\rm exp})]^2/(\Delta M)^2,
\eqno(31)$$
where $E(\mu)$ is obtained from  one of the three curves,
$E({\rm exp})$ are the experimental
eigenenergies obtained with the help of (17),
and $\Delta M$ are the experimental errors in the
meson masses, except that we have taken a minimum error
of 1 MeV and increased some errors to take isospin
mass splittings into account.

Other authors, using potential models,  have obtained
similar values of the $B_c^*$ mass. For example,
Martin [\Ma], using a power law potential, obtained
a value of 6318 MeV, and Eichten and Quigg [\QEQ], using
various potentials, obtained values between 6319 and
6343 MeV.
Bagan et al.\ [\Ba] have averaged a variety of other
people's results to obtain $6330\pm 20$.
On the other hand, some authors  have obtained
quite different values of the mass of the $B_c^*$.
For example, Jain and Munczek [\JM] find $M(B_c^*)=
6277$ MeV.

\medskip
\centerline{VII. OTHER MESON MASS PREDICTIONS}
\medskip
As we have already remarked, we do not expect
the pseudoscalar meson energy eigenvalues to be
monotonically decreasing as a function of $\mu$. However,
once we have a prediction for the $B_c^*$ mass, we can
obtain estimates for pseudoscalar meson masses in
other ways. We use semi-empirical
mass formulas [\Song,\WL] for the splitting between
vector and pseudoscalar states. These semi-empirical
formulas are based on the colormagnetic interaction
as given by the Fermi--Breit theory. However, the formulas
take into account the fact that the strong-interaction
coupling constant runs. The formulas also make empirical
corrections which depend on quark masses so as to get
improved agreement with known data compared to the
Fermi--Breit formula. The semi-empirical formulas then
may be used to predict mass splittings in cases
where experimental data are absent.

Following [\WL], we take the hyperfine splitting in mesons
to be given by:
$$M_V-M_S=p \alpha_s(2\mu)\mu^q/(m_1+m_2),\eqno(32)$$
where $m_1$ and $m_2$ are the constituent quark masses,
and we have determined the two free parameters $p$ and $q$
from a fit to the experimental splittings
to be $q=0.642$ and $p=1.917$ (GeV)$^{2-q}$.
The running coupling constant is given by
$$\alpha_s(Q)={4 \pi\over \beta_0t + (\beta_1/\beta_0)\ln t},
\eqno(33)$$ where
$$t=\ln (Q^2/\Lambda_{\rm QCD}^2), \quad \beta_0=11-2n_f/3,
\quad \beta_1=102-38n_f/3,\eqno(34)$$
with $\Lambda_{\rm QCD}=100$ MeV, and $n_f=4$.
We use the quark masses of the first row of Table I
rather than the masses in [\WL], so that our values
of the parameters $q$ and $p$ differ a little from those in
[\WL].

We show in Fig.\ 3 the eigenenergies of the pseudoscalar
mesons, where we use the data [\PDG] on pseudoscalar masses
as input together with the quark masses obtained for the
vector mesons. We omit the $\eta$ and $\eta'$ mesons, because,
as we have already remarked, they are mixed states of
uncertain quark content. We also give in Fig.\ 3 the
eigenenergies of the pseudoscalars obtained from the
corresponding vectors with the aid of
the semi-empirical mass formula (32).
We see from Fig.\ 3 that, except
(as expected) for the pion, the eigenenergies obtained
from the observed masses and from the semi-empirical
formulas are in remarkably good agreement. Note that
the eigenenergies of the light pseudoscalars
violate the condition that $E(\mu)$ be monotonically decreasing.
We expect this violation because the colormagnetic term
for light pseudoscalars gives a large positive
contribution to $\partial E/\partial \mu$.

Using the semi-empirical mass formula, we obtain a splitting
in the $B_c$ system of $65\pm 10$ MeV, and in the $\bar bb$
system of $55\pm 10$ MeV.
We estimate that the masses of the $B_c$ and $\eta_b$ are
$$M(B_c) = 6255\pm 30\ {\rm MeV}, \quad
M(\eta_b) = 9405\pm 15\ {\rm MeV}. \eqno(35)$$
We give the mass of the $B_c$ in Table II.

We next turn to excited vector meson  states, for which
the data are considerably poorer than for the ground
states. Furthermore, complications might arise from
possible mixing with four-quark, hybrid,
and glueball states. For example, the excited
$\rho(1465)$ and $\omega(1394)$ states differ in mass by
about 70 MeV, although they ought to be degenerate in our
model. We believe this difference indicates appreciable
mixing. Similar considerations apply to light $P$-wave
mesons.  Therefore, we use only the charmonium and bottomonium
data of the Particle Data Group [\PDG] and confine ourselves
to excited $B_c^*$ states. Because we have only two data
points for each excited state (a $\psi$ and an $\Upsilon$)
we cannot do better than use a linear fit to predict the
masses of missing states, again using the quark masses in
the first row of Table I. The problem with a linear fit is
that a straight line is not concave upward, so that the
predictions made in this fashion should be regarded as
upper limits. We show in column 4 of Table II our predictions
for the upper limits of two excited vector meson $B_c^*$ states.

Quigg [\Quigg] suggested that it might be better to interpolate
between mass differences, since these are considerably
smaller than the masses themselves. With  this procedure
we have no theoretical reason to reject a linear
interpolation.
We show in column 5 of Table II our predictions for the
masses of  vector $B_c^*$ excited states using linear
interpolation of the mass differences between corresponding
states in the $c\bar c$ and $b\bar b$ systems.
Again the errors include not only statistical errors but
an estimate of the errors associated with the procedures.
Note that the predicted masses in column 5 are less
than the upper limits of column 4, {\it i.e.},
the latter are indeed an upper bound.

Turning to the $P$-wave mesons, we are not able to
show analytically that
the sum of the Fermi--Breit tensor and spin-orbit
interactions  satisfies the inequality (5).
Nevertheless, it turns out empirically that if we
use the same values of quark masses as for the vectors
(row 1 of Table I), the eigenenergies of the
tensor ($J^P=2^+$), axial vector
($J^P=1^+$), and scalar ($J^P=0^+$) mesons
satisfy (9) and (25). We exploit this fact to fit
separate 3-parameter exponential curves to  the
tensors, axial vectors, and scalars so as to obtain
predictions for the $B_c$ $P$-wave states. These are
shown in column 4 of Table II. Our estimated
errors are rather large because of deviations of the
curve from the eigenenergies of the known mesons.

We can also use linear interpolation between $\bar c c$
and $\bar bb$ states to obtain  the masses of $B_c$
$P$-wave states. These are also given in column 5 of
Table II together with estimated errors. We see from
Table II that the predictions for the $P$-wave mesons
in columns 4 and 5 agree within the errors.

\medskip
\centerline{VIII. BARYON MASSES}
\medskip

Because the masses of only four baryons of spin 3/2 are
known experimentally and none of these contains any heavy
quarks, the Feynman--Hellmann theorem {\it by itself}
does not enable us to make useful predictions
of the masses of any baryons containing heavy quarks.
However, if we use the Feynman--Hellmann theorem in
conjunction with a semi-empirical formula for the
colormagnetic splitting in baryons [\WL], we are able
to make some useful estimates of unknown masses.
The reason is that the masses of the $\Lambda_c$
(quark content $qqc$),  $\Sigma_c$ ($qqc$),
$\Xi_c$ ($qsc$), and $\Lambda_b$ ($qqb$) spin-1/2 baryons
are known from experiment [\PDG,\UA1,\OPAL], so that we
can estimate the masses of the corresponding spin-3/2
baryons from a semi-empirical formula for the colormagnetic
splitting in baryons [\WL]. We are then able to use a
procedure analogous to that we used for mesons in order to
obtain estimates of the masses of unknown baryons.

The expression (32) for mesons can be generalized to
baryons as follows [\WL]: We order the quarks so
that if two quarks have the same flavor, they are
chosen to be the first two; if all quark flavors are
different, then the first two are the lightest.
We denote by $M^*$ the mass of the ground state spin-3/2
baryon, with $M_S$ the mass of the ground state spin-1/2
baryon in which the first two quarks have spin 1, and with
$M_A$ the mass of the spin-1/2 baryon whose first two quarks
are in a relative spin 0 state. We take the two-quark
colormagnetic matrix elements as
$$8R_{ij,k}=F_{ij,k}p
\alpha_s(2\mu_{ij})\mu_{ij}^q/(m_i+m_j),
\eqno(36)$$
with
$$F_{ij,k}=[\mu_{ij} +x(\mu_{ik}+\mu_{jk})]/
(\mu_{ij} +\mu_{ik}+\mu_{jk}) \eqno(37)$$
to simulate the shrinking of the wave function with
increasing mass $m_k$ of the spectator quark.
The expression for $F_{ij,k}$
in Eq.\ (37) is slightly different from
that given in Ref.\ [\WL] and fits the data somewhat
better.  Following [\WL],  we write the ground-state
baryon mass differences as:
$$\eqalign {
M^*-M_S= & 3R_{13,2}+3R_{23,1},    \cr
M_S-M_A= & 4R_{12,3}-2R_{13,2}-2R_{23,1}.\cr }\eqno(38)$$
The parameters $ n_f$ and $\Lambda_{QCD}$ are chosen
to be the same as in the meson case, while $p$, $q$, and
$x$ are adjustable.

Although the structure of (38) is motivated by
perturbation theory, we take these semi-empirical mass
formulas to be more generally applicable, the justification
being the good agreement  with observed baryon
mass splittings.

A difficulty is that, unlike in the meson case, where
we determined the quark masses from a fit to the
vector meson eigenenergies,  we do not
know {\it a priori} what input values to use for
quark masses which will be best suited for baryons.
Our procedure is to start with $m_q= 300$ MeV and
the other quark masses taken at reasonable starting
values, for example, with the values given in [\WL]. We
then adjust the quark masses by an iteration procedure
which we shall now describe.

\item{(1)} We use input quark masses and adjust
the parameters $p$, $q$, and $x$ to get a best fit
to the known colormagnetic splittings in baryons. We
then use the semi-empirical mass formula
to calculate the masses of three spin-3/2
baryons ($\Sigma_c^*$, $\Xi_c^*$, and $\Sigma_b^*$)
which are not known from experiment.

\item{(2)} We then use the Feynman--Hellmann theorem,
analogously to the meson case; i.e., we adjust the
parameters of a three-parameter curve and the quark
masses $m_s$, $m_c$ and $m_b$ to get a best fit
to the eigenenergies of the $\Delta$, $\Sigma^*$,
$\Xi^*$, $\Omega$, $\Sigma_c^*$,
$\Xi_c^*$, and $\Sigma_b^*$.

\item{(3)} We then use the new quark masses in the
semi-empirical mass formula and repeat steps (1) and (2).

In practice, this method rapidly converges. We find
the best parameters  of the semi-empirical
mass formula (38) are $p=0.331$ (GeV)$^{2-q}$, $q=0.417$,
$x=3.805$ when used with the following
(rounded) quark masses for baryons in MeV:
$$m_q=300, \quad m_s=475, \quad m_c=1640, \quad m_b=4990.
\eqno(39)$$
The mass difference $m_s-m_q$ satisfies the inequality (21).
The parameters of an exponential curve of form (26) turn
out to be
$$a=1307 \ {\rm MeV}, \quad b=757 \ {\rm MeV}, \quad
c=813 \ {\rm MeV}. \eqno(40)$$

There are several reasons why our
procedure for baryons is not as precise as that
we used for mesons. First, the input ``data'' for baryons
include three baryon  masses which do not come from
experiment but are only estimated from a mass
formula. Second, even using these three baryons,
we have only seven baryons to obtain the three
parameters of a curve and the three quark masses.
Third, we have to obtain unknown baryon masses by
extrapolation, which is a less precise method than
the interpolation method used for mesons.

Comparing the masses of Eq.\ (39) with those used
for mesons (see the first row of Table I), we see
that the quark  masses which give a best
fit to the  baryons are (except for
$m_q$, which was assumed to be the same)
a little higher than those which lead to a best fit
to the mesons. Because these quark masses
are constituent masses, {\it i.e. effective} ones,
there are no theoretical reasons why the masses
determined from the baryons should coincide exactly
with those determined from the mesons.
If we insist that a single set of
quark masses hold for both baryons and mesons, and
vary these masses,  our overall best fit
to the hadron data is significantly poorer and
our predictions have greater errors.

Using the baryon data only, we can predict the masses
of as yet unobserved baryons from the
Feynman--Hellmann theorem and the baryon semi-empirical mass
formulas. As we have remarked, the baryon masses,
given in Table III,
are obtained by extrapolation, rather than interpolation,
so that the errors are larger than in the meson case.
The errors in the masses of the $\Xi_b$ ($qsb$,
spin 1/2, antisymmetric
in $qs$), $\Xi_b'$ (spin 1/2, symmetric in $qs$),
$\Xi_b^*$ (spin 3/2), $\Omega_c$ ($ssc$, spin 1/2),
$\Omega_c^*$ (spin 3/2), $\Omega_b$ ($ssb$, spin 1/2), and
$\Omega_b^*$ (spin 3/2)
arise partly because of the
substantial error in the measurements to date [\UA1,\OPAL]
of the mass of the $\Lambda_b$.
We believe that the following predicted mass differences
(in MeV) are likely to have smaller errors
than any of the masses given in Table III.
$$M(\Sigma_b)-M(\Lambda_b) =200\pm 20,\quad
M(\Sigma_b^*)-M(\Lambda_b) =230\pm 20, $$
$$M(\Xi_b)-M(\Lambda_b)= 190\pm 30,  \quad
M(\Xi^{\prime}_b)-M(\Lambda_b)= 330\pm 30,
\eqno(41)$$
$$M(\Xi_b^*)-M(\Lambda_b)= 360\pm 30.$$
It is interesting that our semi-empirical mass formula
makes the $\Sigma_b$ about 10 MeV heavier than the
$\Xi_b$. However, the probable error is such that
this is not a firm prediction. All we can really say
is that the $\Sigma_b$ and $\Xi_b$ have masses which
are very likely within 20 MeV of each other.

\medskip
\centerline{IX. CONCLUSIONS}
\medskip

In conclusion, we have shown that the
Feynman--Hellmann theorem leads to the  inequality
$\delta E/\delta \mu < 0$ for most
ground-state vector mesons and spin 3/2 baryons, even
in the presence of relativistic kinematics and a
flavor-dependent colormagnetic interaction.
We were not able to show this for
hadron pairs which differ by one member of the pair
containing both a heavier and a lighter quark
than the other, but the result seems to be empirically
true even in this case. This inequality and
the concavity condition (25), provide theoretical
justification for an interpolation method [\KR], which
allows one to make a quantitative prediction about the
mass of the $B_c^*$ and other mesons without assuming
any specific functional form for the quark-antiquark
interaction. We obtain the masses of still other mesons
containing heavy quarks by using a semi-empirical mass
formula and by interpolating among mass differences.
For the baryons, we can also use (8) and (25) to obtain
predictions, but we need the semi-empirical mass formula
from the outset and also need to extrapolate, rather than
interpolate, in order to obtain useful results. Therefore,
our predictions are not as precise as in the meson case.
In making  our predictions of the values
of heavy hadron masses, we have not had to assume
an explicit form for the Hamiltonian, but only some
general characteristics about its flavor dependence.

\medskip
\centerline{ACKNOWLEDGMENTS}
\medskip

We should like to thank Boris Kopeliovich and Malcolm
Macfarlane for valuable discussions and Chris
Quigg for a helpful comment. This work was
supported in part by the U. S. Department of Energy
and in part by the U. S. National Science Foundation.
\vfill\eject

References
\bigskip

\item{[\QR]} C. Quigg and J. L. Rosner, Phys. Rep. {\bf 56},
167 (1979). %1

\item{[\LSG]} W. Lucha, F. F. Sch\"oberl, D. Gromes,
Phys. Rep.  {\bf 200}, 127 (1991). %2

\item{[\MNSMM]} S. N. Mukherjee, R. Nag, S. Sanyal, T. Mori,
J. Morishita, M. Tsuge, Phys. Rep. {\bf 231}, 201 (1993). %3

\item{[\Ri]} J.-M. Richard, Phys. Rep. {\bf 212}, 1 (1992). %4

\item{[\DBL]} D. B. Lichtenberg,  Int.\ J. Mod.\ Phys. A
{\bf 2}, 1669 (1987). %5

\item{[\Di]} B. Dieckmann, Phys. Rep. {\bf 159}, 99 (1988). %6

\item{[\FS]} D. Flamm and F. Sch\"oberl
{\it Introduction to the quark model of elementary
particles}, (Gordon and Breach, New York, 1982). %7

\item{[\GR]} S. Gasiorowicz and J. L. Rosner, Am.\ J.\
Phys.\ {\bf 49}, 954 (1981). %8

\item{[\KR]} W. Kwong and J. L. Rosner, Phys.\ Rev.\ D
{\bf 44}, 212 (1991). %9

\item{[\CL]}  I. Cohen and H. Lipkin, Phys.\ Lett.\
{\bf 84B}, 323 (1979);
H. Lipkin, Phys.\ Lett.\ {\bf B 319}, 276 (1993). %10

\item{[\Fe]} R.~P. Feynman, Phys. Rev. {\bf 56}, 340 (1939).%11

\item{[\He]} H. Hellmann, Acta Physicochimica URSS
I, 6, 913 (1935); IV, 2, 225 (1936); Einf\"uhrung
in die Quantenchemie (F. Deuticke,
Leipzig and Vienna, 1937) p. 286. %12

\item{[\GM]} H. Grosse and A. Martin, Phys. Rep.
{\bf 60}, 341 (1980). %13

\item{[\Song]} X. Song, Phys.\ Rev.\ D
{\bf 40}, 3655 (1989). %14

\item{[\WL]} Yong Wang and D. B. Lichtenberg,
Phys.\  Rev.\ D {\bf 42}, 2404 (1990). %15

\item{[\don]} D. B. Lichtenberg,  Phys.\ Rev.\ D {\bf 40},
4196 (1989). %16

\item{[\Ei]} E. Eichten {\it et al.}, Phys.\ Rev.\ D
{\bf 21}, 203 (1980). %17

\item{[\GI]} S. Godfrey and N. Isgur, Phys.\ Rev.\
D {\bf 32},  189 (1985). %18

\item{[\BBHM]} W. H. Blask, U. Bohm, M. G. Huber,
B. Ch. Metzsch, and H. R. Petry, Z. Phys.\ A {\bf 337},
327 (1990) and references therein. %19

\item{[\DL]} D. B. Lichtenberg, Phys.\ Rev.\ D {\bf 40},
3675 (1989). %20

\item{[\BO]}R.~A. Bertlmann and S. Ono,
Phys. Lett. {\bf 96B}, 123 (1980).  %21

\item{[\PDG]} Particle Data Group, K. Hikasa {\it et al.},
Phys.\ Rev.\ D {\bf 45}, S1 (1992). %22

\item{[\CDF]} CDF Collaboration, F. Abe {\it et al.},
Phys.\ Rev.\ Lett.\ {\bf 71}, 1685 (1993). %23

\item{[\ALEPH]} ALEPH Collaboration, D. Buskulic
{\it et al.}, Phys.\ Lett.\ B, {\bf 311}, 425 (1993). %24

\item{[\Ono]} S. Ono, Phys.\ Rev.\ D {\bf 17}, 888 (1978).%25

\item{[\BCN]} R. K. Bhaduri, L. E. Cohler, and
Y. Nogami, Nuovo Cimento {\bf 65A}, 376 (1981). %26

\item{[\SB]} B. Silvestre-Brac, Phys.\ Rev.\ D {\bf 46},
2179 (1992). %27

\item{[\Fu]} L. P. Fulcher (unpublished). %28

\item{[\M]} A. Martin, Phys.\ Lett.\ {\bf 93B}, 338 (1980).%29

\item{[\Ma]} A. Martin, in {\it Heavy Flavours and
High-Energy Collisions in the 1--100 TeV Range},
edited by A. Ali and L. Cifarelli, (Plenum Press,
New York, 1989), p. 141. %30

\item{[\BM]} R.~A. Bertlmann and A. Martin,
Nucl. Phys. B {\bf 168}, 111 (1980). %31

\item{[\Th]} W. Thirring, {\it A Course in Mathematical
Physics}, {\bf Vol. 3} (Springer Verlag, Berlin, 1981). %32

\item{[\QEQ]} C. Quigg, Fermilab-Conf-93/265-T
(unpublished); E. Eichten and C. Quigg, Fermilab-pub-94/032-T
(1994, unpublished). %33

\item{[\Ba]} E. Bagan, H. G. Dosch, P. Gosdzinsky, S. Narison,
and J.-M. Richard, CERN-TH.7141/94 (1994, unpublished). %34

\item{[\JM]} P. Jain and H. Munczek,
Phys.\ Rev.\ D {\bf 48}, 5403 (1993).  %35

\item{[\Quigg]} C. Quigg, private communication. %36

\item{[\UA1]} UA1 Collaboration, C. Albajar {\it et al.},
Phys. Lett. {\bf B 273}, 540 (1991). %37

\item{[\OPAL]} OPAL Collaboration, Marco Cattaneo,
Beauty and Charm Hadronic Decays at LEP, at
{\it Advanced Study Conference on Heavy Flavours}, Sept.
3--7, 1993, Pavia, Italy. This is a preliminary measurement.%38

\vfill\eject

TABLE I. Values of quark constituent masses in MeV for
calculating meson energy eigenvalues from
experimental values of their masses. We show in the
first row the (rounded off) values of the quark masses
used in this work and, for comparison, values used by
some other authors in subsequent rows.

\vskip 12pt
$$\vbox {\halign {\hfil #\hfil &&\quad \hfil #\hfil \cr
\cr \noalign{\hrule}%\cr
\cr \noalign{\hrule}
\cr
Reference & $m_q$ & $m_s$ & $m_c$ & $m_b$\cr
\cr \noalign{\hrule}
\cr
 This work & 300 & 440 & 1590 & 4920 \cr
 [\KR]$^*$ & 263 & 404 & 1543 & 4876 \cr     %Kwong, Rosner
 [\WL] & 300 & 500 & 1800 & 5200 \cr         %Wang and DBL
 [\GI] & 220 & 419 & 1628 & 4977 \cr         %Godfrey, Isgur
 [\BO] & 310 & 620 & 1910 & 5270 \cr         %Bertlmann, Ono
 [\Ono]  & 336 & 510 & 1680 & 5000 \cr       %Ono
 [\BCN]  & 337 & 600 & 1870 & 5259 \cr       %Bhaduri et al.
 [\FS]   & 350 & 500 & 1500 & 4700 \cr       %Flamm and Schoberl
{}~[\Song]$^*$ & 270 & 600 & 1700 & 5000 \cr   %Song
 [\Ei] & 335 & 450 & 1840 &5170 \cr          %Eichten
 [\SB]  & 330 & 550 & 1650 & 4715 \cr        %Silvestre-Brac
 [\Fu] & 150 & 366 & 1320 & 4749 \cr         %Fulcher
%[\RT]  & 300 & 600 & 1895 & 5255 \cr        %Richard, Taxil PL 128B (83) 453
\cr \noalign{\hrule}%\cr
\cr \noalign{\hrule}
}}$$
$^*$One of several sets of quark masses in this reference.

\vskip 2cm
TABLE II. Predicted masses of as yet unobserved
$B_c$ ($\bar b c$) mesons. In column 4 we show predictions
for the ground-state vector  and $P$-wave mesons
and upper limits for two excited
vectors from interpolation of the energy eigenvalues,
using the Feynman--Hellmann theorem. In column 5 we
show the pseudoscalar mass obtained from  a semi-empirical
mass formula, and excited vector and $P$-wave states
from interpolation of mass differences.

\vskip 12pt
$$\vbox {\halign {\hfil #\hfil &&\quad \hfil #\hfil \cr
\cr \noalign{\hrule}%\cr
\cr \noalign{\hrule}
\cr
Name &Spin-parity $J^P$ & $n~^{2S+1}L_J$ & Mass (MeV) &
Mass (MeV) \cr
\cr \noalign{\hrule}
\cr
$B_c^*$ & $1^-$ & $1~^3 S_1$ & $6320\pm 20$ &    \cr
$B_c$   & $0^-$ & $1~^1 S_0$ &  & $6255\pm 30$    \cr
%$D_s^*$ & $1^-$ & $2~^3S_1$  & $2620\pm 30$ & $2700\pm 40$ \cr
$B_c^*$ & $1^-$ & $2~^3 S_1$ & $<6940$ & $6900\pm 20$ \cr
$B_c^*$ & $1^-$ & $3~^3 S_1$ & $<7290$ & $7250\pm 20$ \cr
$B_c^*$ & $0^+$ & $1~^3 P_0$ & $6630\pm 40$ & $6660\pm 30$ \cr
$B_c^*$ & $1^+$ & $1~^3 P_1$ & $6730\pm 40$ & $6740\pm 30$ \cr
$B_c^*$ & $2^+$ & $1~^3 P_2$ & $6770\pm 40$ & $6780\pm 30$ \cr
%$B_c^*$ & $\bar bc$ & $1~^1 P_1$ &  & $6750\pm 30$ \cr
\cr \noalign{\hrule}%\cr
\cr \noalign{\hrule}
}}$$
\vfill\eject

TABLE III. Predicted masses of as yet unobserved
baryons. In column 3 we show a prediction
for a ground-state spin-1/2 baryon ($\Xi_b$)
whose first two quarks have an antisymmetric spin wave
function. (Antisymmetric $\Omega_c$ and $\Omega_b$
states do not exist in our picture.)
Column 4 shows predictions for ground-state spin-1/2 baryons
with symmetric spin wave function in the first two quarks.
In column 5 we show predictions for the ground-state
spin-3/2 baryons. See Eq.\ (41) for (we believe, more reliable)
estimates for baryon mass differences.

\vskip 12pt
$$\vbox {\halign {\hfil #\hfil &&\quad \hfil #\hfil \cr
\cr \noalign{\hrule}%\cr
\cr \noalign{\hrule}
\cr
Name & Quark content & $M_A$ (MeV) & $M_S$ (MeV) &
$M^*$ (MeV) \cr
\cr \noalign{\hrule}
\cr
$\Lambda_b, \Sigma_b, \Sigma_b^*$ & $qqb$ &
$^{\dagger}5630\pm 40$ & $5830\pm 40$ & $5860\pm 40$ \cr
$\Xi_b,\Xi_b',\Xi_b^*$ &$qsb$ & $~5820\pm 40$ &
$5960\pm 40$ & $5990\pm 40$ \cr
$\Omega_c, \Omega_c^*$ &$ssc$ & ---    & $2710\pm 50$ &
$2770\pm 50$ \cr
$\Omega_b, \Omega_b^*$ & $ssb$ & ---   & $6070\pm 60$ &
$6100\pm 60$ \cr
\cr \noalign{\hrule}%\cr
\cr \noalign{\hrule}
}}$$
$^{\dagger}$Input

\vskip 2cm

Figure captions
\bigskip

FIG.\ 1. Energy eigenvalues of vector mesons using experimental
masses from Refs.\ [\PDG--\ALEPH] in conjunction with four
sets of quark masses from Table I:
(a) from Ref. [\BO], (b) from Ref. [\GI], (c) from Ref. [\FS],
and (d) from Ref. [\SB].
In the figure, the letters $a$ through $i$ stand for
$\rho$, $K^*$, $\phi$, $D^*$, $B^*$, $D_s^*$, $B_s^*$,
$J/\psi$, and $\Upsilon$ respectively.

\medskip

FIG.\ 2. Energy eigenvalues of vector mesons using our
quark masses from the first row of Table I and the
experimental masses from Refs.\ [\PDG--\ALEPH].
The letters stand for the same mesons as in Fig. 1,
and the solid circle is our prediction for the $B_c^*$.
The solid line is a fit to the vector meson data with an
exponential form, Eq.\ (26), with parameters given in Eq.\ (29).

\medskip

FIG.\ 3. Eigenenergies of the pseudoscalar mesons obtained
from the masses of the Particle Data Group [\PDG]  (crosses)
compared with the eigenenergies from the vector meson
masses and the semi-empirical mass formula of Eq.\ 32) (open
circles). The quark masses of the first row of Table I were
used to obtain eigenergies from masses. In order of increasing
$\mu$ are $\pi$, $K$, $D$, $B$, $D_s$, $B_s$,  $\eta_c$,
$B_c$, and $\eta_b$.

\bye